\documentclass[fleqn,usenatbib,useAMS]{mnras}
 \usepackage{graphicx}
 \usepackage{amsmath}
 \usepackage{amssymb}
 \usepackage[T1]{fontenc}
 \usepackage{ae,aecompl}
 \usepackage{txfonts}
 \usepackage{enumerate}
 \title[Twisted extreme trans-Neptunian orbital space]
       {Twisted extreme trans-Neptunian orbital parameter space: 
        statistically significant asymmetries confirmed}

 \author[C. de la Fuente Marcos and R. de la Fuente Marcos]
        {C.~de~la~Fuente~Marcos$^{1}$\thanks{E-mail: nbplanet@ucm.es}
         and
         R. de la Fuente Marcos$^{2}$ \\
         $^1$Universidad Complutense de Madrid,
             Ciudad Universitaria, E-28040 Madrid, Spain \\
         $^2$AEGORA Research Group,
             Facultad de Ciencias Matem\'aticas,
             Universidad Complutense de Madrid,
             Ciudad Universitaria, E-28040 Madrid, Spain}
 \date{Accepted 2022 February 3. 
       Received 2022 February 1; 
       in original form 2021 December 15}
 \pubyear{2022}
 
\begin{document}
  \label{firstpage}
  \pagerange{\pageref{firstpage}--\pageref{lastpage}}
  \maketitle
%
%
  \begin{abstract}
     Asymmetric debris discs have been found around stars other than the 
     Sun; asymmetries are sometimes attributed to perturbations induced by 
     unseen planets. The presence or absence of asymmetries in our own 
     trans-Neptunian belt remains controversial. The study of sensitive 
     tracers in a sample of objects relatively free from the perturbations 
     exerted by the four known giant planets and most stellar flybys may 
     put an end to this debate. The analysis of the distribution of the 
     mutual nodal distances of the known extreme trans-Neptunian objects 
     (ETNOs) that measure how close two orbits may get to each other could 
     be such a game changer. Here, we use a sample of 51~ETNOs together 
     with random shufflings of this sample and two unbiased scattered-disc 
     orbital models to confirm a statistically significant (62$\sigma$) 
     asymmetry between the shortest mutual ascending and descending nodal 
     distances as well as the existence of multiple highly improbably 
     ($p<0.0002$) correlated pairs of orbits with mutual nodal distances as 
     low as 0.2~au at 152~au from the Solar system's barycentre or 1.3~au 
     at 339~au. We conclude that these findings fit best with the notion 
     that trans-Plutonian planets exist.
  \end{abstract}

  \begin{keywords}
     methods: statistical -- celestial mechanics --
     minor planets, asteroids: general -- Kuiper belt: general --
     Oort Cloud.
  \end{keywords}

  \section{Introduction}
     Asymmetric debris discs or exoKuiper belts are sometimes the result of perturbations induced by companions (see e.g. 
     \citealt{2020MNRAS.498.1319M,2021AJ....161..271F,2021MNRAS.506.1978L}). When no stellar or substellar companions are detected, the 
     observed asymmetries are often attributed to secular interactions with unseen planets (see e.g. \citealt{2019A&A...631A.141S}, but see
     also \citet{2021AJ....162..278Z}). In this respect, our local trans-Neptunian belt is not too different from the so-called exoKuiper 
     belts (see e.g. \citealt{2018ARA&A..56..541H, 2020tnss.book..351W}), but our vantage point does not help in confirming or rejecting the 
     presence of asymmetries out there. The existence of asymmetries or gaps in our local trans-Neptunian belt has been suggested (see e.g. 
     \citealt{2017AJ....154...62V,2020A&A...637A..87L,2021AJ....162...39O}) or rejected (see e.g. \citealt{2019AJ....158...49V,
     2021ARA&A..59..203G}) based on nearly the same input data and similar diagnostic tools, particularly the statistical analysis of the 
     orbital inclination distributions. In order to produce robust results, the identification of asymmetries in the trans-Neptunian or 
     Kuiper belt requires the study of sensitive tracers in a sample of objects the least possible affected by perturbations from the four 
     known giant planets and most stellar flybys. 

     A promising instance of tracer and object sample is in the distribution of mutual nodal distances of the extreme trans-Neptunian 
     objects (ETNOs) ---TNOs with semimajor axis, $a>150$~au and perihelion distance, $q>30$~au--- as discussed by 
     \citet{2021MNRAS.506..633D}. Using data for 39 ETNOs, \citet{2021MNRAS.506..633D} found that the distribution of mutual nodal distances 
     of the studied sample showed a statistically significant asymmetry between the shortest mutual ascending and descending nodal distances, 
     hereinafter the ${\Delta}_{+}$--${\Delta}_{-}$ asymmetry. In addition, a highly improbable pair of ETNOs made of (505478) 
     2013~UT$_{15}$ and 2016~SG$_{58}$ with a mutual ascending nodal distance of 1.35~au at 339~au from the barycentre of the Solar system 
     was uncovered by the analysis; the probability of finding such a pair by chance was estimated to be 0.000205$\pm$0.000005. 

     The Dark Energy Survey (DES) has recently announced the discovery of 815 new TNOs \citep{2022ApJS..258...41B}. After this data release, 
     the ETNO data set now includes 51~objects (an increase of nearly 31 per cent, from 39) and the ${\Delta}_{+}$--${\Delta}_{-}$ 
     distribution has 1275 pairs instead of 741 (a 72 per cent increase). Here, we use this larger data set to further investigate the 
     ${\Delta}_{+}$--${\Delta}_{-}$ asymmetry and the presence of improbable orbit pairs. This Letter is organized as follows. In Section~2, 
     we discuss data and methods. Theoretical expectations from two different models are reviewed in Section~3. The updated distribution of 
     mutual nodal distances is presented and assessed in Section~4. Section~5 shows that the statistically significant asymmetries found by 
     \citet{2021MNRAS.506..633D} are confirmed. Our results are discussed in Section~6 and our conclusions are summarized in Section~7.

  \section{Data and methods}
     The distribution of the mutual nodal distances of the known ETNOs provides information on how close two orbits may get to each other. 
     The shortest mutual nodal distance is the minimum orbit intersection distance or MOID that, in absence of protective mechanisms such as 
     mean-motion or secular resonances, could be close to the minimum approach distance between two ETNOs; in other words, a sufficiently 
     short nodal distance might be signalling recurrent close flybys and, perhaps, even a past physical connection between two ETNOs. 
     Here, we compute the distribution of the mutual nodal distances of the known ETNOs (51 as of 2-February-2022) as described in 
     appendix~A of \citet{2021MNRAS.506..633D}, using eqs.~16 and 17 in \citet{2017CeMDA.129..329S}.

     For a given pair of objects, the computation of the mutual nodal distances requires the values of semimajor axis, $a$, eccentricity, 
     $e$, inclination, $i$, longitude of the ascending node, $\Omega$, and argument of perihelion, $\omega$. As in 
     \citet{2021MNRAS.506..633D}, we have used barycentric values accessed from Jet Propulsion Laboratory's (JPL) Horizons on-line solar 
     system data and ephemeris computation service\footnote{\href{https://ssd.jpl.nasa.gov/?horizons}{https://ssd.jpl.nasa.gov/?horizons}} 
     that uses the new DE440/441 general-purpose planetary solution \citep{2021AJ....161..105P}. Data were queried using tools provided by 
     the \textsc{Python} package \textsc{Astroquery} \citep{2019AJ....157...98G}. Statistical analyses were performed using \textsc{NumPy} 
     \citep{2011CSE....13b..22V,2020Natur.585..357H} and visualized using the \textsc{Matplotlib} library \citep{2007CSE.....9...90H}.

  \section{Theoretical expectations}
     Synthetic ETNOs can be randomly generated in accordance with a nominal scattered-disc model. The study of the distribution of nodal 
     distances of sets of simulated objects can provide an unbiased reference to evaluate the statistical significance of any features 
     identified in the nodal distances distribution of real ETNOs. Following the analysis in \citet{2021MNRAS.506..633D}, we have considered 
     the scattered-disc models discussed by \citet{2001AJ....121.2804B,2017AJ....154...65B} and \citet{2021PSJ.....2...59N}. The single 
     difference between both models is in the ($a$, $e$, $q$) set of parameters. While \citet{2021PSJ.....2...59N} favours a model in which 
     $a$ follows the distribution $N(a) \propto a^{0.7}$ and $e$ is uniformly distributed in the interval (0.69, 0.999), 
     \citet{2001AJ....121.2804B,2017AJ....154...65B} argues for a uniform distribution in both $a$ and $q$. As described in appendix~C of 
     \citet{2021MNRAS.506..633D} and for the latter scattered disc model, $a$ follows a uniform distribution in the interval (150, 1000)~au, 
     $q$ is also uniform in the interval (30, 100)~au, the angular elements $\Omega$ and $\omega$ are drawn from a uniform distribution in 
     the interval (0\degr, 360\degr), and $i$ follows the so-called Brown distribution of inclinations 
     \citep{2001AJ....121.2804B,2017AJ....154...65B} in the interval (0\degr, 60\degr). 

\begin{figure}
       \centering
        \includegraphics[width=\linewidth]{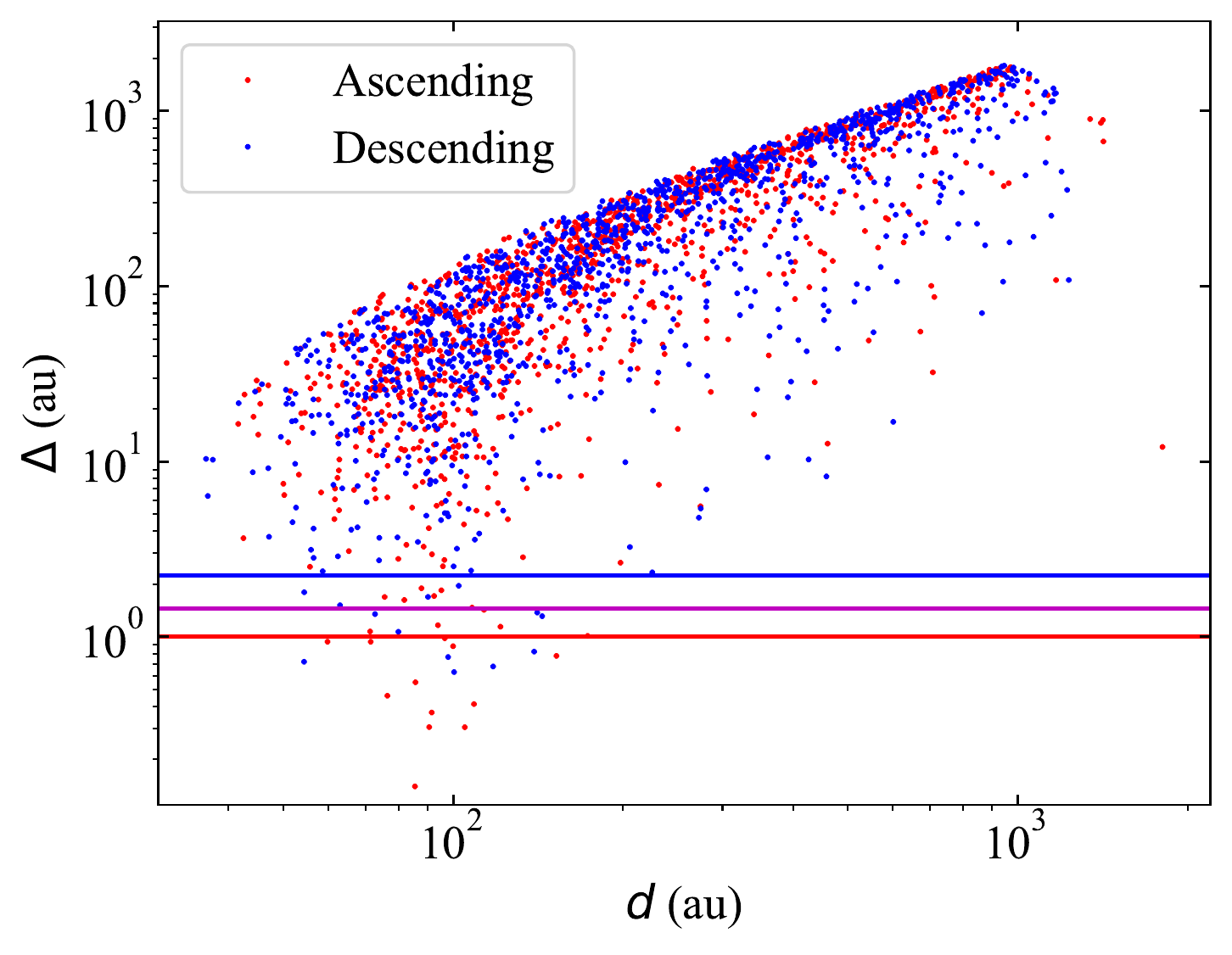}
        \includegraphics[width=\linewidth]{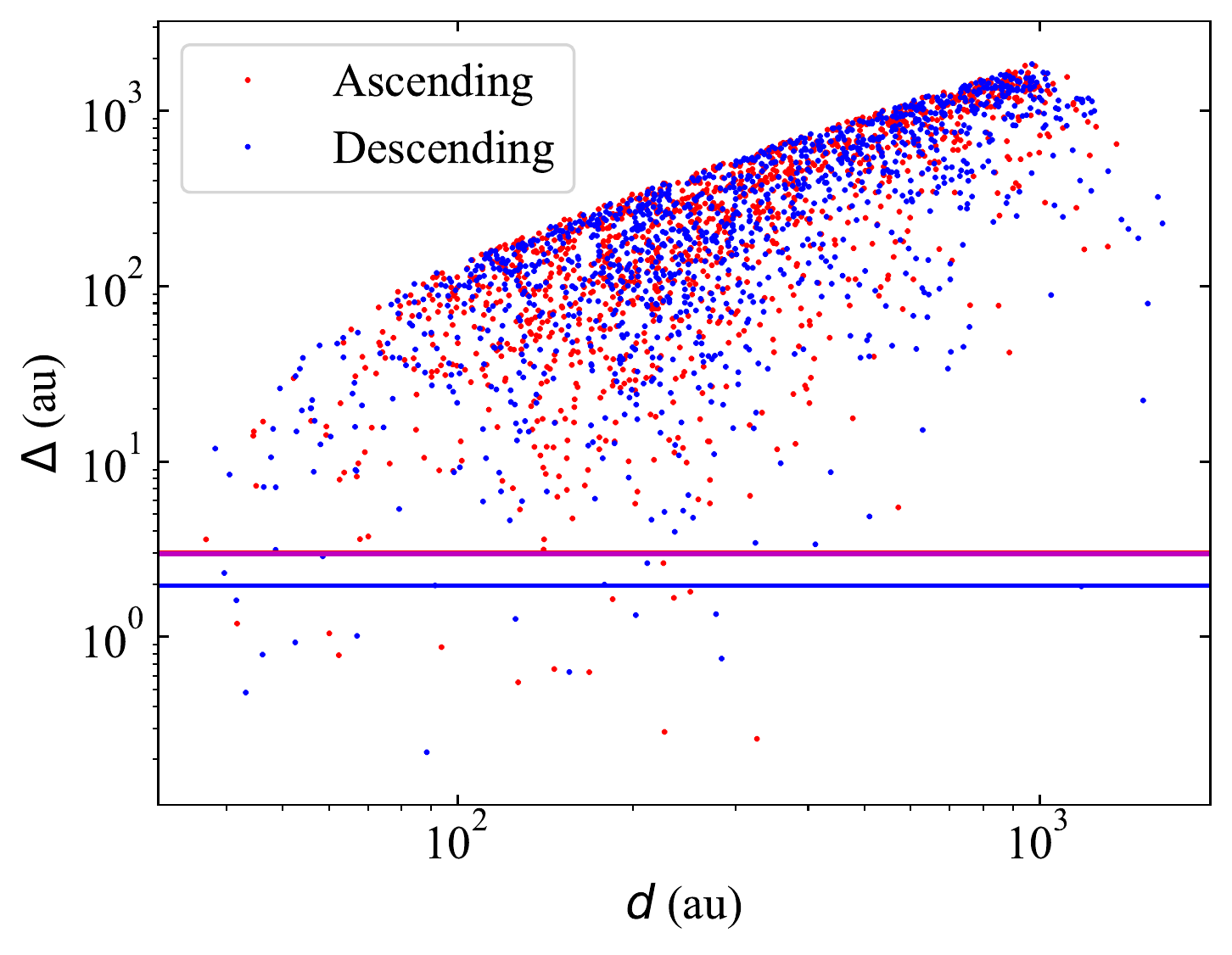}
        \caption{Mutual nodal distances as a function of the barycentric distance to the node for a representative sample of 51 synthetic 
                 ETNOs or 1275 pairs. The top panel shows results compatible with the scattered-disc model discussed by 
                 \citet{2001AJ....121.2804B,2017AJ....154...65B}; the bottom panel displays results consistent with the model in 
                 \citet{2021PSJ.....2...59N}. The values of mutual nodal distances of ascending nodes, ${\Delta}_{+}$, are shown in red and 
                 those of descending nodes, ${\Delta}_{-}$, in blue. The horizontal lines show the relevant first percentiles. In purple, we 
                 show 1.455~au (top) and 2.975~au (bottom). See the text for details. 
                }
        \label{nodaldistsynt}
     \end{figure}

     Figure~\ref{nodaldistsynt} shows representative distributions of both scattered-disc models corresponding to 51 synthetic ETNOs. In 
     \citet{2021MNRAS.506..633D} and after analyzing 10 instances of 10$^{4}$ synthetic ETNOs each (or 49995000 pairs per instance) randomly 
     generated in accordance with \citet{2001AJ....121.2804B,2017AJ....154...65B}, it was found that the first percentile (that signals 
     extreme outliers) of ${\Delta}_{+}$ was 1.460$\pm$0.009~au and the one for ${\Delta}_{-}$ was 1.450$\pm$0.009~au, which implies that, 
     under the assumptions in this model, the median and 16th and 84th percentiles of the first percentiles of the mutual ascending and 
     descending nodal distances must be identical. A similar result was found for the model in \citet{2021PSJ.....2...59N}, but in this case 
     the first percentile of ${\Delta}_{+}$ is 2.98~au and the one for ${\Delta}_{-}$ is 2.97~au. Using a steeper slope for the semimajor 
     axis distribution (e.g. $\propto a^{1.3}$) further increases the value of the extreme outliers threshold. In the following, we will 
     consider 1.455~au as a reference for the first percentile value of ${\Delta}_{\pm}$. 
%
%
%
%

  \section{Updated nodal distances distribution}
\begin{figure}
       \centering
        \includegraphics[width=\linewidth]{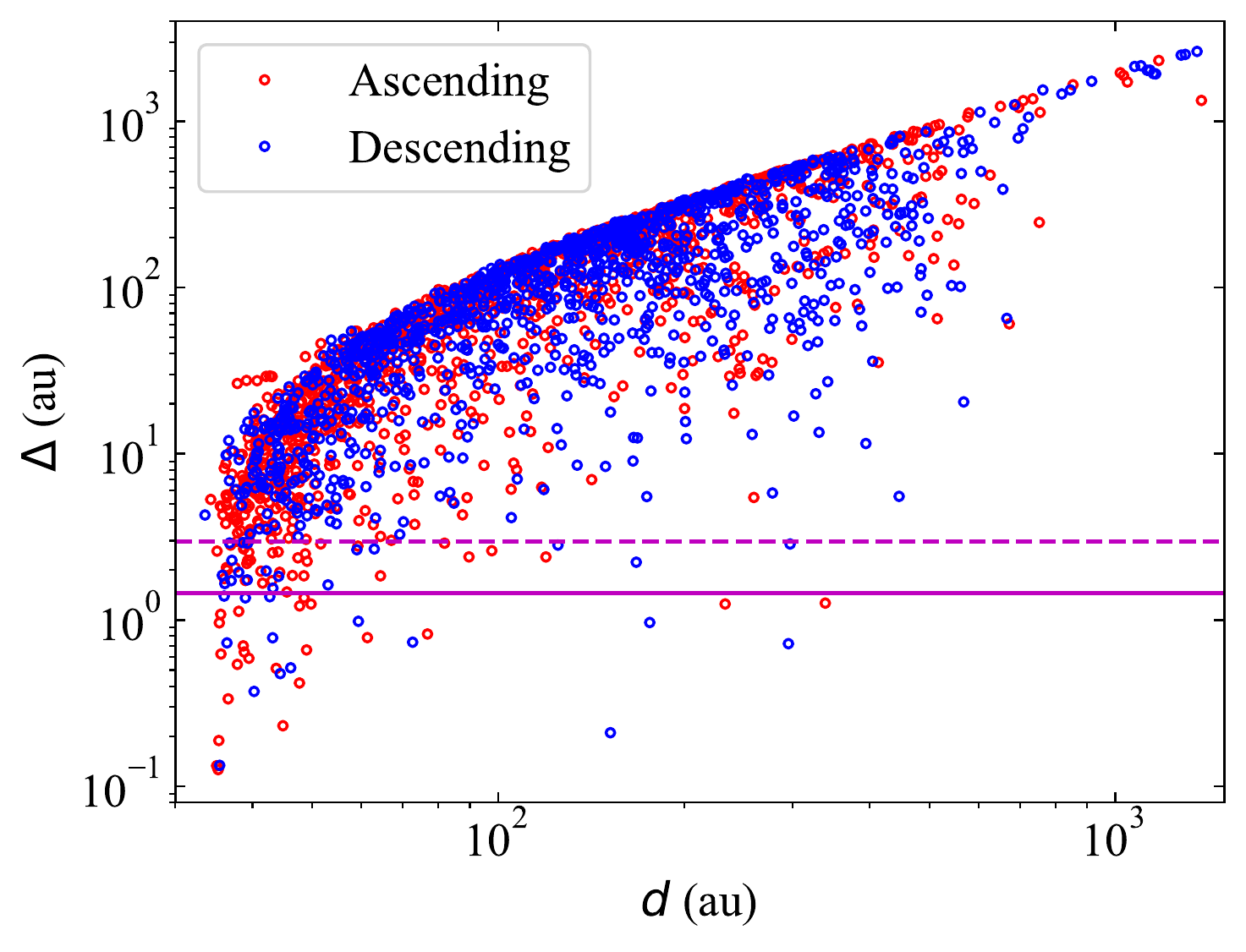}
        \caption{Mutual nodal distances as a function of the barycentric distance to the node for the sample of 51 ETNOs or 1275 pairs. The
                 values of mutual nodal distances of ascending nodes, ${\Delta}_{+}$, are shown in red and those of descending nodes,
                 ${\Delta}_{-}$, in blue. The solid purple line corresponds to 1.455~au and the dashed one to 2.975~au. See the text for 
                 details.  
                }
        \label{nodaldist}
     \end{figure}

     The nodal distances distribution corresponding to the 51 known ETNOs or 1275 pairs was computed as described in appendix~A of 
     \citet{2021MNRAS.506..633D}, taking into account both the nominal (mean) values of the barycentric orbital elements and their 
     uncertainties. Figure~\ref{nodaldist} shows the median values of the mutual nodal distances as a function of the average barycentric 
     distance to the node. The values of mutual nodal distances of ascending nodes, ${\Delta}_{+}$, are displayed in red and those of 
     descending nodes, ${\Delta}_{-}$, in blue. As in \citet{2021MNRAS.506..633D} and in order to identify severe outliers, we use the first 
     percentile of the distribution (see e.g. \citealt{2012psa..book.....W}) that is 0.768$\pm$0.004~au for ${\Delta}_{+}$ and 
     1.392$\pm$0.010~au for ${\Delta}_{-}$ (means and standard deviations from 10 experiments). 
%
%
%
%

     We have tested the significance of the correlations found above by shuffling the orbital elements: the Fisher--Yates shuffle 
     \citep{FiYa1938,Knuth1969} in the form implemented in \textsc{Python}'s \textsc{random.shuffle} was applied to each orbital element 
     independently to produce a random permutation of the original sequence. Each randomized data set (sampling 51 orbits $10^{6}$ times) 
     was processed as the original one was. When considering the nominal values of the orbital elements ${\Delta}_{\pm}$=1.353~au; if both 
     mean values and uncertainties are included in the shuffling process, then ${\Delta}_{\pm}$=1.454~au. This randomization test shows that 
     by removing any existing correlations in the data set, the ${\Delta}_{+}$--${\Delta}_{-}$ asymmetry is also removed. We interpret this 
     result as signalling that the asymmetry is intrinsic to the data set and not an artifact or due to chance. On the other hand, 
     ${\Delta}_{\pm}$=1.454~au is close to the value ${\Delta}_{-}$=1.392$\pm$0.010~au (but statistically inconsistent, 6.2$\sigma$); we 
     interpret this as evidence in favour of the mutual descending nodes being rather randomly distributed while the distribution of mutual 
     ascending nodes not being statistically compatible (171.5$\sigma$) with a random one. 

     On the other hand, \citet{2021MNRAS.506..633D} found one highly improbably correlated pair of orbits; with the new data, we found five 
     pairs of ETNOs with mutual nodal distances below the first percentile of the distribution and barycentric distance above 150~au. The 
     five pairs are (${\Delta}_{\pm}$ as median value and 16th and 84th percentiles): (527603) 2007~VJ$_{305}$ and 2015~VQ$_{207}$ with 
     ${\Delta}_{-}=0.21^{+0.21}_{-0.15}$~au at 152~au, (594337) 2016~QU$_{89}$ and 2016~SA$_{59}$ with 
     ${\Delta}_{-}=0.72^{+0.72}_{-0.51}$~au at 295~au, (496315) 2013~GP$_{136}$ and 2014~SX$_{403}$ with 
     ${\Delta}_{-}=0.97^{+1.03}_{-0.68}$~au at 176~au, 527603 and 2013~RC$_{156}$ with
     ${\Delta}_{+}=1.25^{+1.36}_{-0.88}$~au at 233~au, and (505478) 2013~UT$_{15}$ and 2016~SG$_{58}$ with 
     ${\Delta}_{+}=1.26^{+1.31}_{-0.89}$~au at 339~au. As for the statistical significance of these highly improbably correlated pairs 
     within the context of the randomized data set, the probability of having mutual nodal distances below the first percentile of the 
     distribution at a barycentric distance above 300~au is 0.00014507$\pm$0.00000013 and that of having mutual nodes closer than 0.25~au 
     beyond 150~au is 0.00013444$\pm$00000004. It is therefore confirmed that within the known ETNOs data set, the five pairs pointed out 
     above are indeed remarkable.

  \section{Statistically significant asymmetries confirmed}
     In Section~3, we have shown that ${\Delta}_{+}=0.768\pm0.004$~au and ${\Delta}_{-}=1.392\pm0.010$~au. The respective values obtained by 
     \citet{2021MNRAS.506..633D} were 1.450$\pm$0.010~au and 2.335$\pm$0.014~au (for 39 ETNOs or 741 pairs and also from 10 experiments). 
     Therefore, the new data confirm that the distribution of mutual nodal distances has a statistically significant asymmetry between the 
     shortest mutual ascending and descending nodal distances, this time at the 62$\sigma$ level (63$\sigma$ in \citealt{2021MNRAS.506..633D}).

     On the other hand, \citet{2021MNRAS.506..633D} found that the peculiar pair of ETNOs made of (505478) 2013~UT$_{15}$ and 2016~SG$_{58}$ 
     had a mutual ascending nodal distance of 1.35~au at 339~au from the Sun. With the new data, 505478 and 2016~SG$_{58}$ have 
     ${\Delta}_{+}=1.26^{+1.31}_{-0.89}$~au at 339~au and we found four new pairs of ETNOs with mutual nodal distances within the first 
     percentile of the distribution and barycentric distance above 150~au. Within the context of the randomized data set discussed above, 
     these five pairs of ETNOs are highly improbably correlated as the probability of having mutual nodal distances below the first 
     percentile of the distribution at a barycentric distance above 300~au is as low as 0.00014507$\pm$0.00000013.

     In Fig.~\ref{nodaldistsynt}, both models fail to reproduce the high density of large nodal distances observed at barycentric distances 
     below 100~au in Fig.~\ref{nodaldist}, although the model in \citet{2001AJ....121.2804B,2017AJ....154...65B} appears to produce better 
     results. The actual value of ${\Delta}_{-}$ for the sample of 51 known ETNOs is 1.392$\pm$0.010~au that is somewhat consistent with the 
     one from the scattered-disc model discussed by \citet{2001AJ....121.2804B,2017AJ....154...65B}, but incompatible with that of 
     \citet{2021PSJ.....2...59N}. 

     As for the statistical significance of ETNO pairs like the ones pointed out above within the context of the scattered-disc model 
     discussed by \citet{2001AJ....121.2804B,2017AJ....154...65B}, the analysis of $10^{5}$ experiments sampling 51 synthetic orbits each 
     indicates that the probability of finding five pairs with ${\Delta}_{\pm}<1.455$~au at a barycentric distance above 150~au is 
     0.0577$\pm$0.0013 that suggests that the five ETNO pairs could indeed be unusual, although the evidence may not be fully conclusive as 
     there is nearly 6 per cent chance of obtaining a result at least as extreme as the one observed using the scattered-disc model 
     discussed by \citet{2001AJ....121.2804B,2017AJ....154...65B}.

  \section{Discussion}
     The larger ETNO data set assembled thanks to the recent DES data release \citep{2022ApJS..258...41B} has led to significant improvements 
     in the results initially reported by \citet{2021MNRAS.506..633D}. Although it is difficult to argue against the reality of the 
     ${\Delta}_{+}$--${\Delta}_{-}$ asymmetry and the number of strongly correlated orbits seems to defy predictions from two scattered-disc 
     models, the source of these features is still unclear. 

     Following, e.g., \citet{2021PSJ.....2...59N} one may argue that observational bias could be the source of the features observed here;
     in particular, if the components of some or all the peculiar pairs of ETNOs pointed out above were discovered by the same survey or at 
     similar locations in the sky, observational bias could not be easily rejected as irrelevant. Table~\ref{disco} shows the discovery 
     circumstances of the members of the pairs discussed above. No components of a given pair were found by the same survey and only one 
     pair had both components appearing closely projected at discovery time, (594337) 2016~QU$_{89}$ and 2016~SA$_{59}$. Observational bias 
     may not be the source of the highly improbably correlated pairs.
%
%
     \begin{table}
        \centering
        \fontsize{8}{11pt}\selectfont
        \tabcolsep 0.08truecm
        \caption{Discovery circumstances of the members of the pairs discussed in the text. Data include: discovery date, right ascension 
                 ($\alpha$) and declination ($\delta$) at discovery time, heliocentric distance ($r_{\rm d}$) and apparent motion 
                 ($\mu_{\rm d}$) at discovery time, and discovery observatory ---O means OSSOS discovery, A means Apache Point, D means 
                 DECam, and M means Mauna Kea. Data sources: JPL's SSDG SBDB and MPC.
                }
        \begin{tabular}{lrrrrrr}
          \hline
           Object          & disc. date     &
                             $\alpha$                                    & $\delta$                         &
                             $r_{\rm d}$      & $\mu_{\rm d}$                     & obs.         \\ 
                           &                &
                                      ($^{\rm h}$:$^{\rm m}$:$^{\rm s}$) &          (\degr:\arcmin:\arcsec) &
                                              (au) &          (\arcsec\ h$^{-1}$) &              \\ 
          \hline
           496315          & 2013-Apr-04    &
                             14:07:32.88                                 & $-$11:09:38.8                    &
                             45.4                  & 2.63                         &            O  \\
           505478          & 2013-Oct-31    &
                             00:54:20.28                                 &    05:12:29.1                    &
                             61.2                  & 1.91                         &            O  \\
           527603          & 2007-Nov-04    &
                             00:29:31.74                                 & $-$00:45:45.0                    &
                             35.2                  & 2.60                         &            A  \\
           594337          & 2016-Aug-25    &
                             02:44:19.76                                 & $-$00:38:53.5                    &
                             35.5                  & 1.20                         &            D  \\
           2013~RC$_{156}$ & 2013-Sep-04    &
                             20:02:16.73                                 & $-$51:57:29.2                    &
                             38.6                  & 2.46                         &            D  \\
           2014~SX$_{403}$ & 2014-Sep-25    &
                             00:44:30.34                                 & $-$23:49:43.1                    &
                             41.1                  & 3.12                         &            D  \\
           2015~VQ$_{207}$ & 2015-Nov-05    &
                             00:40:57.35                                 & $-$24:29:07.9                    &
                             31.4                  & 2.70                         &            D  \\
           2016~SA$_{59}$  & 2016-Sep-28    &
                             02:41:32.23                                 & $-$00:17:00.2                    &
                             40.7                  & 2.52                         &            M  \\
           2016~SG$_{58}$  & 2016-Sep-27    &
                             02:35:41.45                                 &    01:05:18.7                    &
                             36.1                  & 2.70                         &            M  \\
          \hline
        \end{tabular}
        \label{disco}
     \end{table}
%
%

     On the other hand, objects with $q>30$~au may still experience close interactions with Neptune that perhaps could induce the observed 
     features. A more selective criterion to avoid this issue is to make the cut at larger $q$ (e.g. 37~au or 40~au). Considering $q>40$~au, 
     there are 21 ETNOs or 210 pairs and a similar analysis gives ${\Delta}_{+}=0.578\pm0.004$~au and ${\Delta}_{-}=3.29\pm0.03$~au, which 
     confirms a statistically significant (90$\sigma$) asymmetry. On the other hand, four out of nine objects (44 per cent) in 
     Table~\ref{disco} have perihelion distance $q>37$~au; two out of nine (22 per cent) have $q>40$~au. However, none of the pairs have 
     both members with $q>37$~au.

\begin{figure}
       \centering
        \includegraphics[width=\linewidth]{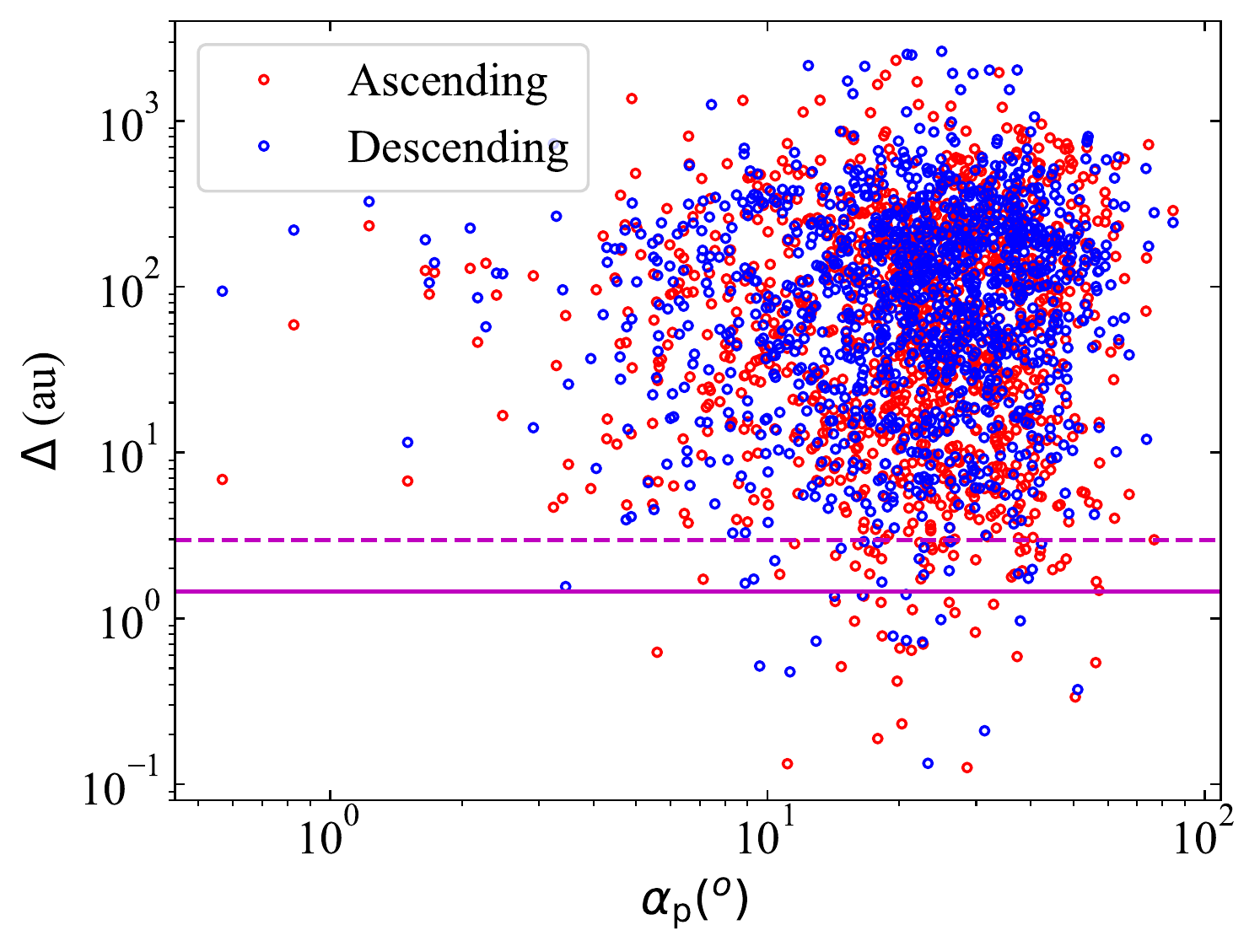}
        \includegraphics[width=\linewidth]{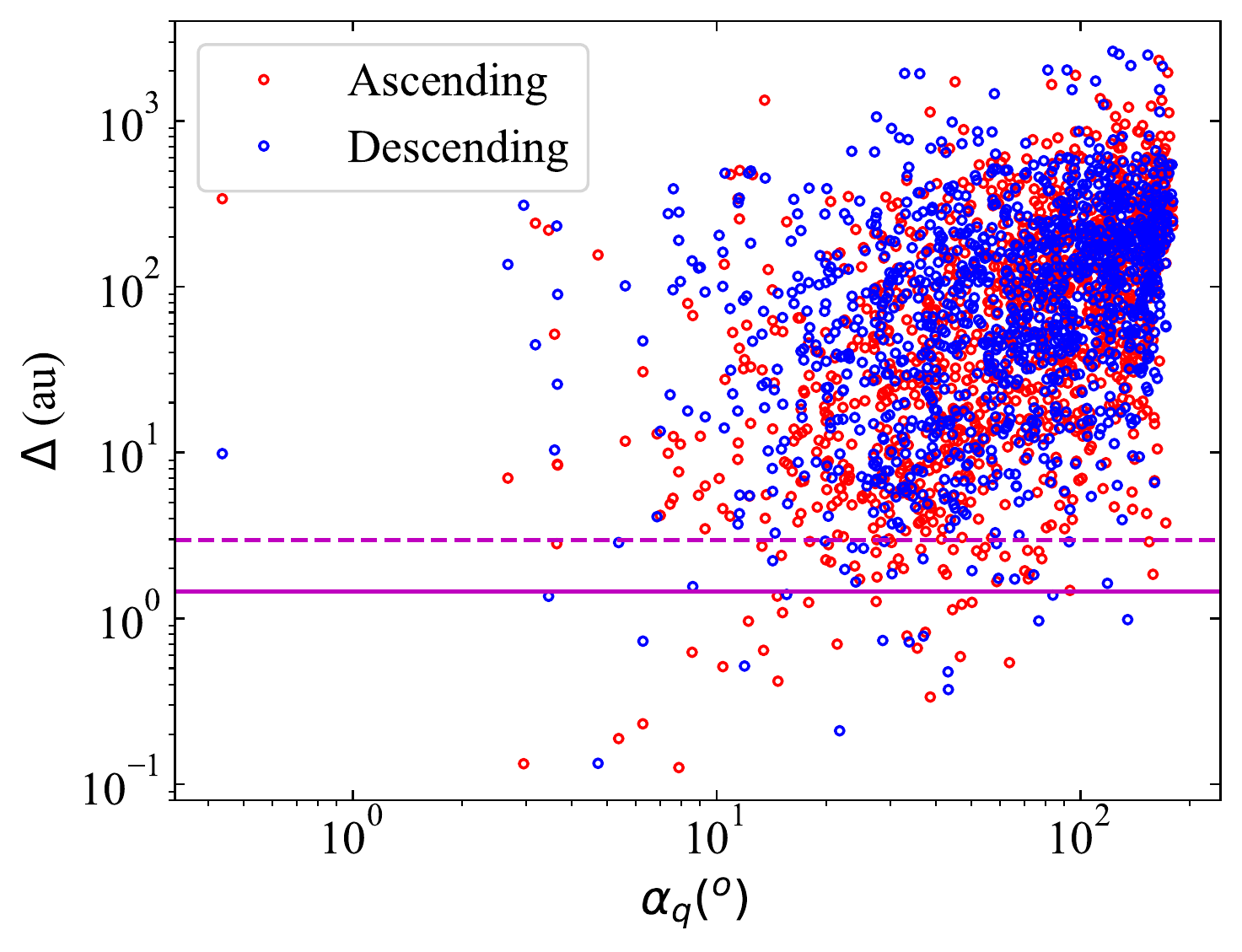}
        \caption{Mutual nodal distances as a function of the angular separations between orbital poles, $\alpha_{\rm p}$ (top panel), and 
                 perihelia, $\alpha_q$ (bottom panel), for the sample of 51 ETNOs or 1275 pairs. The values of mutual nodal distances of 
                 ascending nodes, ${\Delta}_{+}$, are shown in red and those of descending nodes, ${\Delta}_{-}$, in blue. The solid purple 
                 line corresponds to 1.455~au and the dashed one to 2.975~au. See the text for details.
                }
        \label{nodesvsspaceorientation}
     \end{figure}

     Other than bias and the influence of Neptune, perhaps the simplest conjecture with which one may try to account for the observed 
     features is in assuming that they are linked to the presence of families, analogues of the well-studied asteroid families present in 
     the main asteroid belt. The subject of finding collisional families of TNOs has been explored by \citet{2003EM&P...92...49C} and 
     \citet{2011ApJ...733...40M}. Collisional asteroid families have already been found in the regular trans-Neptunian space: the first 
     confirmed family found in the outer Solar system was the one associated with dwarf planet Haumea \citep{2007Natur.446..294B} although 
     a candidate collisional family had previously been proposed by \citet{2002ApJ...573L..65C} and later confirmed by 
     \citet{2018MNRAS.474..838D}, who found four new candidate collisional families of TNOs and a number of unbound TNOs that may have a 
     common origin. However, arguments against the collisional ETNO family conjecture come from different lines of reasoning. On the one 
     hand, there is only one object in common between the five pairs of ETNOs with highly improbably correlated orbits, (527603) 
     2007~VJ$_{305}$, that is a member of two pairs but with nodes at rather different barycentric distances. On the other hand, 
     Fig.~\ref{nodesvsspaceorientation} shows that the unusual pairs have values of the angular separation between orbital poles, 
     $\alpha_{\rm p}$, that are generally above 10{\degr} with the one of pair (505478) 2013~UT$_{15}$--2016~SG$_{58}$ slightly above 
     14{\degr} and that of pair (594337) 2016~QU$_{89}$--2016~SA$_{59}$ being about 23{\degr}, the others are higher; the values of the 
     angular separations between perihelia, $\alpha_q$, are also relatively large. In sharp contrast, the values of $\alpha_{\rm p}$ and 
     $\alpha_q$ are expected to be small when objects come from a recent fragmentation event (see e.g. \citealt{2018MNRAS.474..838D}). The 
     values of $\alpha_{\rm p}$ and $\alpha_q$ were computed as described in appendix~B of \citet{2021MNRAS.506..633D}.
%
%
%
%

     The orbital pole is the intersection between the celestial sphere and a hypothetical axis perpendicular to the plane of the orbit under 
     study. The direction of perihelia is the intersection between the celestial sphere and a hypothetical line that goes from the focus of 
     the orbit towards the point where the orbit under study reaches perihelion. Using observational data, \citet{2018MNRAS.474..838D} 
     showed that any fragments produced by a relatively recent disruption event must have low values of $\alpha_{\rm p}$ and $\alpha_q$, 
     probably under $\sim$2{\degr}; fragments from an old disruption episode may have values of $\alpha_q$ uniformly distributed in the 
     interval (0{\degr},~180{\degr}) and a wide range in the values of $\alpha_{\rm p}$. The distribution of $\alpha_q$ (histogram not 
     shown) is far from uniform. The pairs with low values of $\alpha_{\rm p}$ and $\alpha_q$ in Fig.~\ref{nodesvsspaceorientation} all have 
     mutual nodal distances well above the first percentile of the distribution. All this evidence points away from a collisional origin for 
     the highly correlated ETNOs.

     The fact remains that some pairs of ETNOs have one improbably close mutual nodal distance and that a mechanism must exist to help 
     keeping them that close for a period of time long enough to make them identifiable even in relatively small samples as the one studied 
     here. Close encounters with a massive planetary perturber may play that role. Small mutual nodal distances have been found for a number
     of members of the 29P/Schwassmann-Wachmann~1 comet complex; in this case, nodes of Centaurs are kept close by Jupiter 
     \citep{2021A&A...649A..85D}. Variations of this scenario within the context of the ETNOs have been explored by 
     \citet{2017MNRAS.467L..66D} and \citet{2017Ap&SS.362..198D}. However, the barycentric distances of the five relevant pairs singled out 
     above show a significant dispersion, so either one planetary perturber moves in an eccentric orbit (e.g. $a$$\sim$250~au and 
     $e$$\sim$0.4) or multiple perturbers might be necessary to produce the observed effects. The unlikely presence (due to its 
     intrinsically low probability) of a triplet among the outlier pairs might also be signalling the influence of trans-Plutonian 
     perturber(s). It is unclear if the mechanism proposed by \citet{2021AJ....162..278Z} could be able to produce the observed features; 
     although it leads to a lopsided outer Solar system, it does it via apsidal clustering, which may not be present in the real data (see 
     e.g. \citealt{2021PSJ.....2...59N}). 

  \section{Conclusions}
     Kuiper belt analogues or exoKuiper belts exhibit a diversity of asymmetries and some of them could be caused by unseen planets (see 
     e.g. \citealt{2018ARA&A..56..541H,2020tnss.book..351W}). The question of whether or not our Solar system hosts an asymmetric 
     trans-Neptunian belt has been asked for decades. This Letter argues that, thanks to the new data, we might have finally got a 
     definite answer to this important conundrum. The main conclusions of our study are:
     \begin{enumerate}[(i)]
        \item We confirm the presence of a statistically significant (62$\sigma$) asymmetry between the shortest mutual ascending and
              descending nodal distances in a sample of 51~ETNOs.  
        \item We confirm the existence of five highly improbably correlated pairs of orbits with mutual nodal distances as low as 0.2~au at 
              152~au from the Solar system's barycentre or 1.26~au at 339~au. 
     \end{enumerate}
     We consider that it is improbable that the features discussed here could result from the presence of collisional families, our findings 
     are more supportive of the existence of trans-Plutonian planets.

  \section*{Acknowledgements}
     We thank the referee for constructive and insightful reports, S.~J. Aarseth, J. de Le\'on, J. Licandro, A. Cabrera-Lavers, J.-M. Petit, 
     M.~T. Bannister, D.~P. Whitmire, G. Carraro, E. Costa, D. Fabrycky, A.~V. Tutukov, S. Mashchenko, S. Deen, and J. Higley for comments 
     on ETNOs, A.~B. Chamberlin for helping with the new JPL's Solar System Dynamics website, and A.~I. G\'omez de Castro for providing 
     access to computing facilities. This work was partially supported by the Spanish `Ministerio de Econom\'{\i}a y Competitividad' 
     (MINECO) under grant ESP2017-87813-R and the `Agencia Estatal de Investigaci\'on (Ministerio de Ciencia e Innovaci\'on)' under grant 
     PID2020-116726RB-I00 /AEI/10.13039/501100011033. In preparation of this Letter, we made use of the NASA Astrophysics Data System and 
     the MPC data server.

  \section*{Data Availability}
     The data underlying this paper were accessed from JPL's Horizons (\href{https://ssd.jpl.nasa.gov/?horizons}
     {https://ssd.jpl.nasa.gov/?horizons}). The derived data generated in this research will be shared on reasonable request to the 
     corresponding author.

  \bsp
  \label{lastpage}
\end{document}